%% file: wsdm-submission.tex
\documentclass[sigconf]{acmart}

\makeatletter
\renewcommand\@formatdoi[1]{\ignorespaces} 
\makeatother
\usepackage{balance}

\def\BibTeX{{\rm B\kern-.05em{\sc i\kern-.025em b}\kern-.08emT\kern-.1667em\lower.7ex\hbox{E}\kern-.125emX}}

\usepackage[ruled, noend, noline, linesnumbered]{algorithm2e}

\usepackage{caption}
\usepackage{subcaption}

\usepackage{subfiles}

\usepackage{paralist}
\usepackage{bm}
\usepackage{epstopdf}
\usepackage{xcolor}
\usepackage{framed}
\usepackage{csvsimple}
\usepackage{longtable}
\usepackage[none]{hyphenat}
\usepackage{adjustbox}

\newtheorem{theorem}{Theorem}[section]
\newtheorem{lemma}[theorem]{Lemma}
\newtheorem{claim}[theorem]{Claim}
\newtheorem{corollary}[theorem]{Corollary}

\newcommand{\ignore}[1]{}

\newcommand{\K}{{\cal K}}

\newcommand{\bT}{\boldsymbol{T}}
\newcommand{\bQ}{\boldsymbol{Q}}

\newcommand{\mainalg}{{\sc Pivoter}}
\newcommand{\mainalgrec}{{\sc ExactCliqueCounterRec}}

\newcommand{\Eppstein}{{\sc Eppstein}}
\newcommand{\BKB}{{\sc BKB}}
\newcommand{\BK}{{\sc BK}}
\newcommand{\Tomita}{{\sc Tomita}}

\newcommand{\funcV}{{\sc func-V}}
\newcommand{\funcE}{{\sc func-E}}
\newcommand{\scr}{{\sc SCTBuilder}}
\newcommand{\sct}{\mathop{SCT}}

\newcommand{\Sec}[1]{\hyperref[sec:#1]{\S\ref*{sec:#1}}} 
\newcommand{\Eqn}[1]{\hyperref[eq:#1]{(\ref*{eq:#1})}} 
\newcommand{\Fig}[1]{\hyperref[fig:#1]{Fig.\,\ref*{fig:#1}}} 
\newcommand{\Tab}[1]{\hyperref[tab:#1]{Tab.\,\ref*{tab:#1}}} 
\newcommand{\Thm}[1]{\hyperref[thm:#1]{Theorem\,\ref*{thm:#1}}} 
\newcommand{\Fact}[1]{\hyperref[fact:#1]{Fact\,\ref*{fact:#1}}} 
\newcommand{\Lem}[1]{\hyperref[lem:#1]{Lemma\,\ref*{lem:#1}}} 
\newcommand{\Prop}[1]{\hyperref[prop:#1]{Prop.~\ref*{prop:#1}}} 
\newcommand{\Cor}[1]{\hyperref[cor:#1]{Corollary~\ref*{cor:#1}}} 
\newcommand{\Conj}[1]{\hyperref[conj:#1]{Conjecture~\ref*{conj:#1}}} 
\newcommand{\Def}[1]{\hyperref[def:#1]{Definition~\ref*{def:#1}}} 
\newcommand{\Alg}[1]{\hyperref[alg:#1]{Alg.~\ref*{alg:#1}}} 
\newcommand{\Ex}[1]{\hyperref[ex:#1]{Ex.~\ref*{ex:#1}}} 
\newcommand{\Clm}[1]{\hyperref[clm:#1]{Claim~\ref*{clm:#1}}} 
\newcommand{\Obs}[1]{\hyperref[obs:#1]{Obs.~\ref*{obs:#1}}} 
\newcommand{\Step}[1]{\hyperref[step:#1]{Step~\ref*{step:#1}}} 

\copyrightyear{2020}
\acmYear{2020}
\setcopyright{acmcopyright}
\acmConference[WSDM '20]{The Thirteenth ACM International Conference on Web Search and Data Mining}{February 3--7, 2020}{Houston, TX, USA}
\acmBooktitle{The Thirteenth ACM International Conference on Web Search and Data Mining (WSDM '20), February 3--7, 2020, Houston, TX, USA}
\acmPrice{15.00}
\acmDOI{10.1145/3336191.3371839}
\acmISBN{978-1-4503-6822-3/20/02}

\begin{document}

\fancyhead{}

\sloppy 







\title{The Power of Pivoting for Exact Clique Counting}
%
%
%
%
%

\author{Shweta Jain}
\affiliation{%
  \institution{University of California, Santa Cruz}
  \city{Santa Cruz, CA}
  \country{USA}}
\email{sjain12@ucsc.edu}

\author{C. Seshadhri}
\affiliation{%
  \institution{University of California, Santa Cruz}
  \city{Santa Cruz, CA}
  \country{USA}}
\email{sesh@ucsc.edu}

\begin{abstract}

Clique counting is a fundamental task in network analysis, and even the simplest
setting of $3$-cliques (triangles) has been the center of much recent research. 
Getting the count of $k$-cliques for larger $k$ is algorithmically challenging,
due to the exponential blowup in the search space of large cliques. But a number
of recent applications (especially for community detection or clustering)
use larger clique counts. 
Moreover, one often
desires \emph{local} counts, the number of $k$-cliques per vertex/edge.


Our main result is \mainalg, an algorithm that exactly counts the number
of $k$-cliques, \emph{for all values of $k$}. It is surprisingly effective
in practice, and is able to get clique counts of graphs that were beyond the reach
of previous work. For example, \mainalg{} gets all clique counts in a social network
with a 100M edges within two hours on a commodity machine. Previous parallel
algorithms do not terminate in days. \mainalg{} can also feasibly get local per-vertex
and per-edge $k$-clique counts (for all $k$) for many public data sets with tens
of millions of edges. To the best of our knowledge, this is the first algorithm
that achieves such results.

The main insight is the construction of a Succinct Clique Tree (SCT) that
stores a compressed unique representation of all cliques in an input graph. It is built
using a technique called \emph{pivoting}, a classic approach by Bron-Kerbosch to reduce the 
recursion tree of backtracking algorithms for maximal cliques. Remarkably, the SCT
can be built without actually enumerating all cliques, and provides a succinct
data structure from which exact clique statistics ($k$-clique counts, local counts)
can be read off efficiently.

\end{abstract}


%
%

%
%



\maketitle

\input{intro.tex}

\section{Main Ideas} \label{sec:ideas}
\subfile{ideas}

\subfile{algo}
\section{Experimental results}\label{sec:results}
\subfile{results}

\begin{acks}
  Shweta Jain and C. Seshadhri acknowledge the support of \grantsponsor{}{NSF}{} Awards \grantnum{}{CCF-1740850}, \grantnum{}{CCF-1813165}, and \grantsponsor{}{ARO}{} Award \grantnum{}{W911NF1910294}.
\end{acks}


\scriptsize

\patchcmd{\thebibliography}{\clubpenalty4000}{\clubpenalty10000}{}{}     
\patchcmd{\thebibliography}{\widowpenalty4000}{\widowpenalty10000}{}{}   
\patchcmd{\bibsetup}{\interlinepenalty=5000}{\interlinepenalty=10000}{}{}

\bibliographystyle{ACM-Reference-Format}
\bibliography{wsdm-submission}

\newpage
\end{document}

%% file: intro.tex
\section{Introduction}\label{sec:intro}
Subgraph counting (also known as motif counting, graphlet counting) is a fundamental algorithmic
problem in network analysis, widely applied in domains such as
social network analysis,
bioinformatics, cybersecurity, and physics (refer to tutorial~\cite{SeTi19} and references within).
One of the most important cases is that of \emph{clique counting}. A $k$-clique
is a complete subgraph on $k$ vertices, and has great significance in network analysis
(Chap. 11 of~\cite{HR05} and Chap. 2 of~\cite{J10}).
Indeed, just the special case of $k=3$ (triangle counting) has a rich history in modern
network science. General clique counting has received much attention in recent times
\cite{JhSePi15, MarcusS10,AhNe+15,Escape,JS17,FFF15,DBS18}. There is a line of recent
work on exploiting clique counts for community detection and dense subgraph discovery
\cite{SaSePi14,Ts15,BeGlLe16,TPM17,lu2018community,YiBiKe19}.

%
Despite much effort on this problem, it has been challenging to get scalable algorithms for
clique counting.  
There is a large literature for counting $3$-cliques (triangles) and 
some of these methods have been extended to counting cliques upto size $5$~\cite{JhSePi15, MarcusS10,AhNe+15,Escape}. However, practical algorithms for counting cliques beyond size $5$ have proven to be much harder, and the reason for this is combinatorial explosion. Essentially, as $k$ increases, 
the number of $k$-cliques blows up. 
For large graphs, some recent practical algorithms have succeeded in counting
up to (around) 10-cliques~\cite{FFF15,JS17,DBS18}. They either use randomized approximation
or parallelism to speed up their counting. Besides the obvious problem that they do not scale
for larger $k$, it is difficult to obtain more refined clique counts (such 
as counts for every vertex or every edge).
%
%
\begin{figure*}[t!]

\begin{subfigure}[b]{0.33\textwidth}
    \centering
    \includegraphics[width=\textwidth]{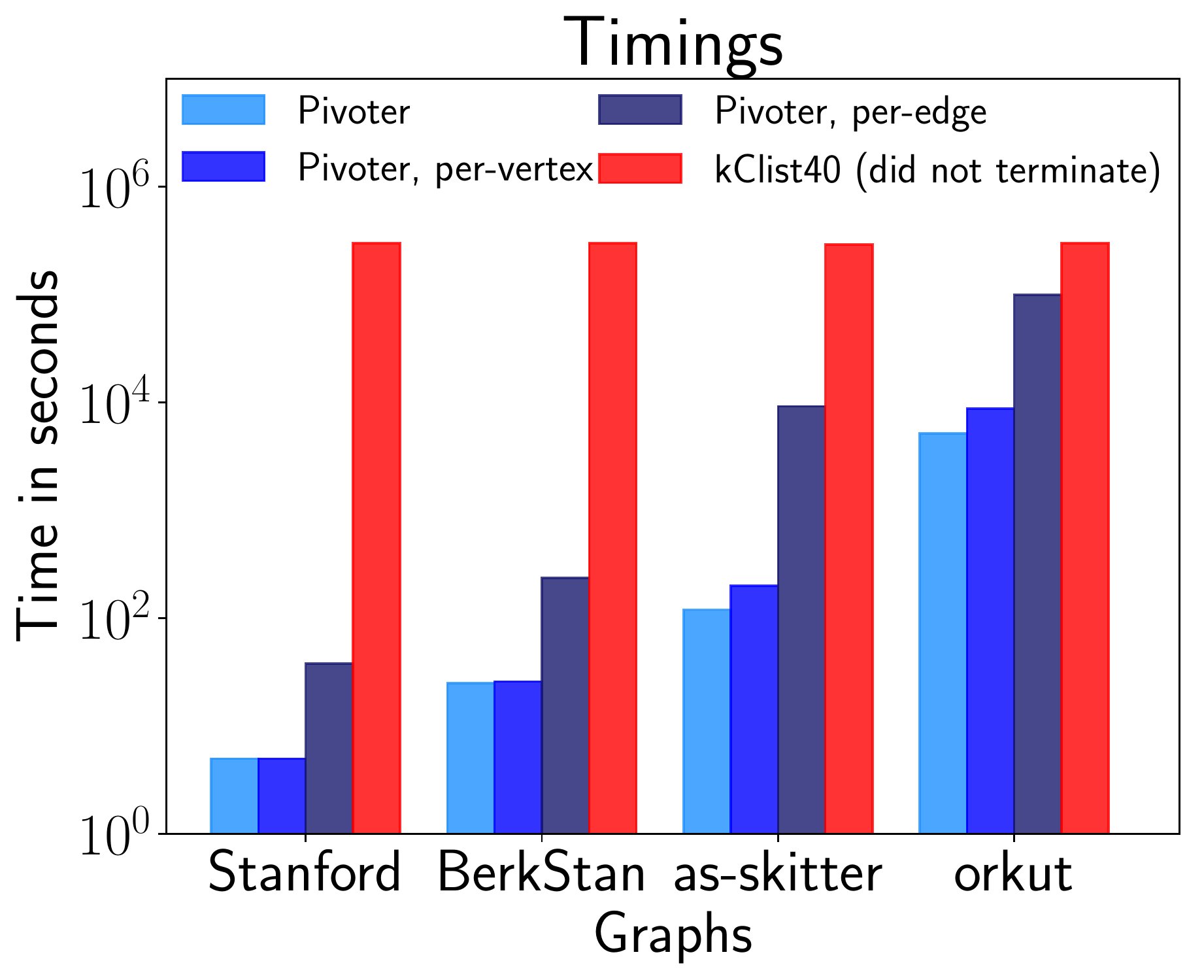}{}
    \caption{\footnotesize{Timings}}
    \label{fig:timings}
    \end{subfigure}
    ~
    \begin{subfigure}[b]{0.33\textwidth}
    \centering
    \includegraphics[width=\textwidth]{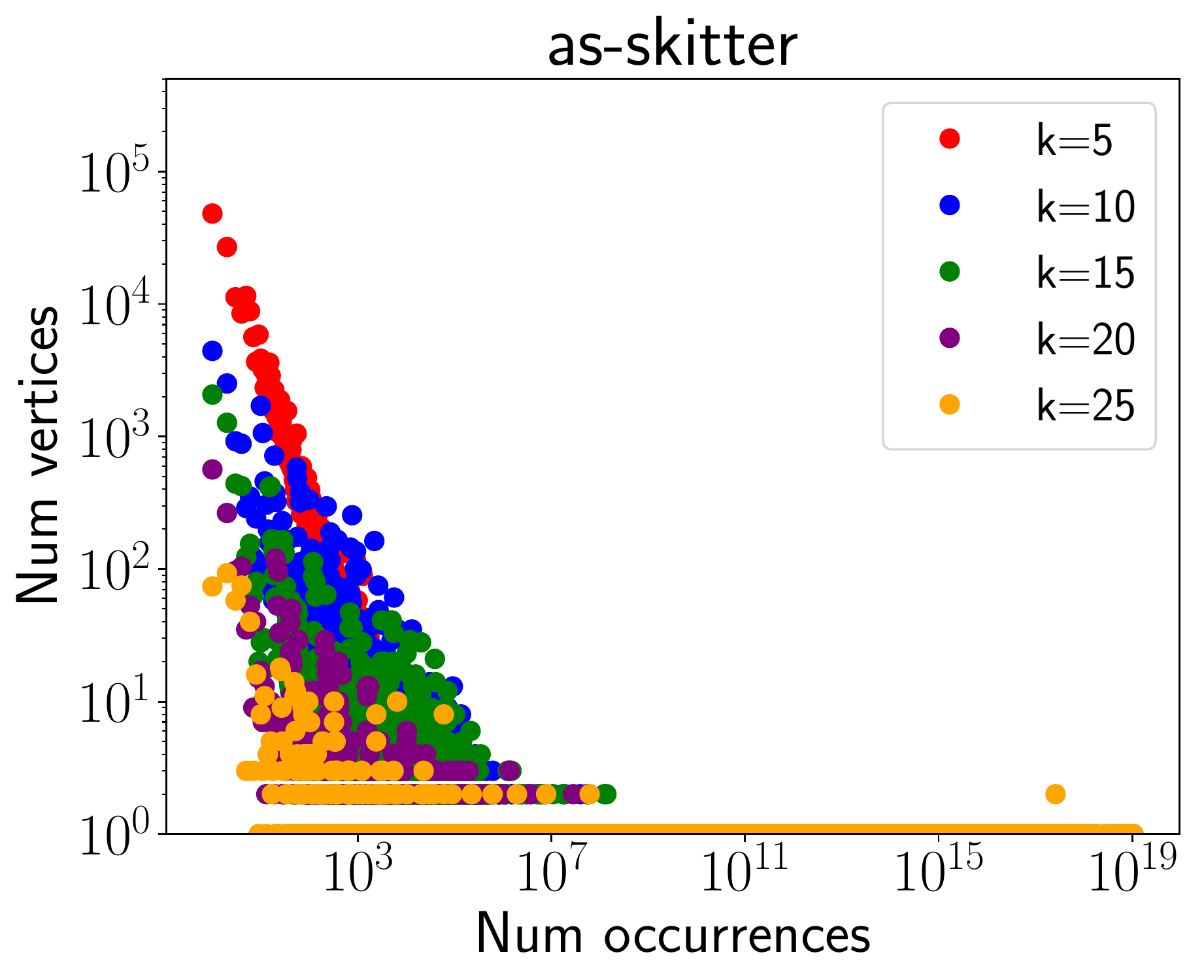}{}
    \caption{\footnotesize{Frequency distribution}}
    \label{fig:soc-pokec-occurrences}
    \end{subfigure}
~
    \begin{subfigure}[b]{0.33\textwidth}
    \centering
    \includegraphics[width=\textwidth]{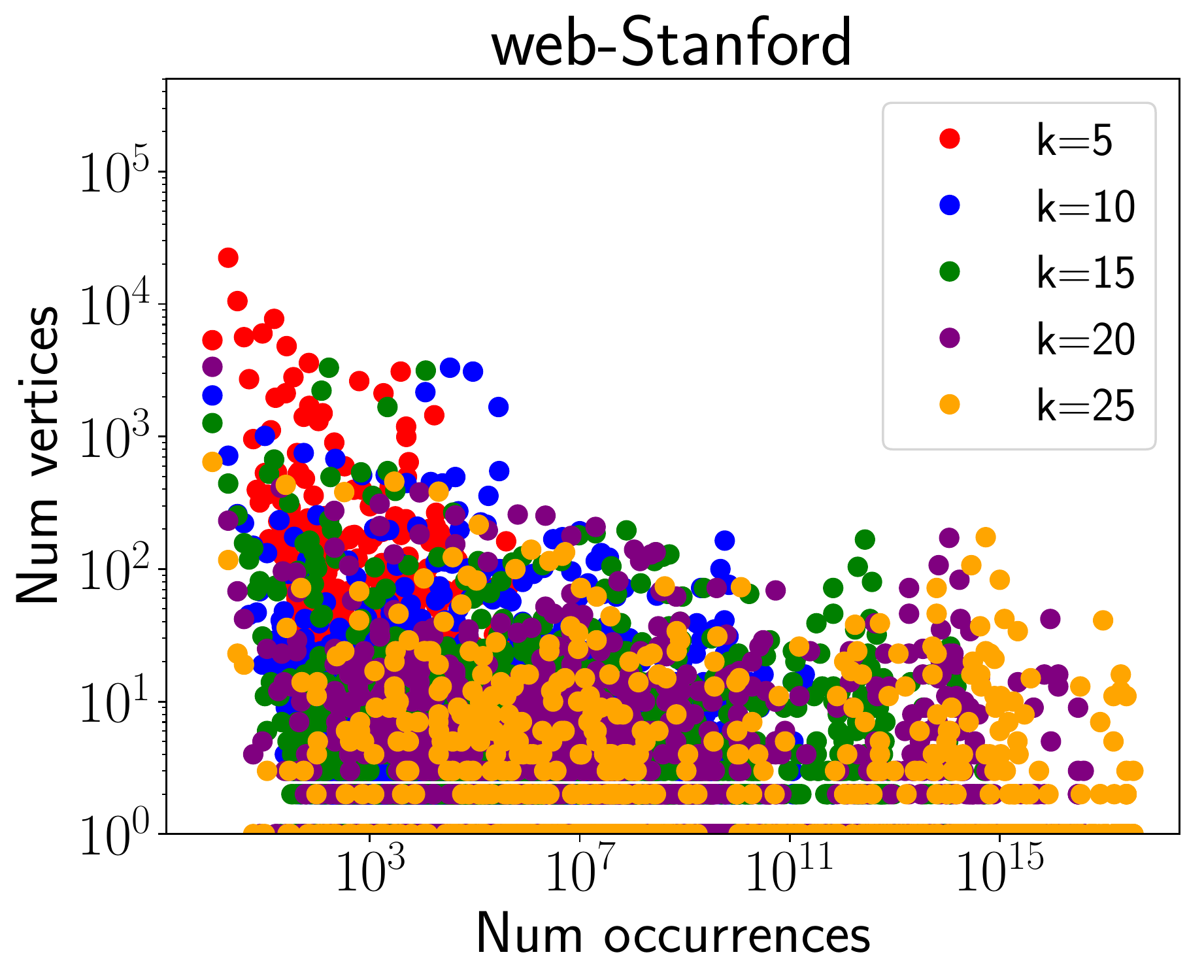}{}
    \caption{\footnotesize{Frequency distribution}}
    \label{fig:web-Stanford-occurrences}
    \end{subfigure}


            \caption{\Fig{timings} shows the comparison of time taken (in seconds) by \mainalg{} for 4 of our largest graphs to count \textit{all} $k-$cliques with the time taken by kClist40 (the parallel version of the state of the art algorithm kClist that uses 40 threads) to count the number of $k-$cliques, where $k$ is the maximum clique size in each graph. For Stanford, BerkStan, as-skitter, orkut, the maximum clique sizes were 61, 201, 67 and 51 resp. \mainalg{} terminated for most graphs in minutes, (except for orkut, for which it took about 2 hours) whereas kClist40 had not terminated even after 3 days, giving a speedup of 100x to 10000x. \Fig{timings} also shows the time taken by \mainalg{} to obtain the per-vertex and per-edge $k-$clique counts. They were within a factor of the time taken to obtain global $k-$clique counts. \Fig{soc-pokec-occurrences} and  \Fig{web-Stanford-occurrences} shows the frequency distribution of $k$-cliques i.e. for every number $r$ on the x-axis, the y-axis shows the number of vertices that participate in $r$ $k$-cliques, for $k \in [5, 10, 15, 20, 25]$ for as-skitter and web-Stanford graphs. } 
\end{figure*}



\subsection{Problem Statement} \label{sec:problem}

We are given an undirected, simple graph $G(V,E)$. For $k \geq 3$, a $k$-clique 
is a set of $k$ vertices that induce a complete subgraph (it contains all edges
among the $k$ vertices). We will denote the number of $k$-cliques as $C_k$.
For a vertex $v \in V$, we use $c_k(v)$ to denote the number of $k$-cliques
that $v$ participates in. Analogously, we define $c_k(e)$ for edge $e \in E$.

We focus on the following problems, in increasing order of difficulty. 
We stress that $k$ is \emph{not} part
of the input, and we want results for all values of $k$.

\begin{asparaitem}
    \item Global clique counts: Output, $\forall k \geq 3$, $C_k$.
    \item Per-vertex clique counts: Output, $\forall k$, $\forall v \in V$, the value $c_k(v)$.
    \item Per-edge clique counts: Output, $\forall k$, $\forall e \in E$, the value $c_k(e)$.
\end{asparaitem}

The per-vertex and per-edge counts are sometimes called \emph{local counts}.
In clustering applications, the local counts
are used as vertex or edge weights, and are therefore even more
useful than global counts~\cite{SaSePi14,Ts15,BeGlLe16,TPM17,lu2018community,YiBiKe19}.

\medskip

\textbf{Challenges:} Even the simplest problem of getting global 
clique counts subsumes a number of recent results on clique counting
\cite{FFF15,JS17,DBS18}. The main challenge
is combinatorial explosion: for example, the {{\tt web-Stanford}} web graph
with 2M edges has \emph{3000 trillion} $15$-cliques. These numbers are even
more astronomical for larger graphs. Any method that tries to enumerate
is doomed to failure. 

Amazingly, recent work by Danisch-Balalau-Sozio
uses parallel algorithms to count beyond trillions of cliques.
But even their algorithm fails to get all global clique counts for a number
of datasets. Randomized methods
have been used with some success, but even they cannot estimate
all clique counts~\cite{FFF15,JS17}.

Local counting, for all $k$, is even harder, especially given the sheer
size of the output. Parallel methods would eventually need to store
local counts for every subproblem, which would increase the overall memory footprint.
For local counts, sampling would require far too many random variables, each of which
need to be sampled many times for convergence.
(We give more explanation in \Sec{related}.)

This raises the main question:

{\em Is there a scalable, exact algorithm for getting all global and local cliques counts,
on real-world graphs with millions of edges?} 

To the best of our knowedge, there is no previous algorithm that can solve these problems
on even moderate-sized graphs with a few million edges. 

%


\subsection{Main contributions} \label{sec:contri}
Our main contribution is a new practical algorithm \mainalg{} for the global and local clique counting
problems. 

\textbf{Exact counting without enumeration:} Current methods for exact clique counting
perform an \emph{enumeration}, in that the algorithm explicitly ``visits" every clique. Thus,
this method cannot scale to counting larger cliques, since the number of cliques is simply too large.
Our main insight is that the method of \emph{pivoting}, used  
to reduce recursion trees for maximal clique 
enumeration~\cite{BK73,ELS13}, can be applied to counting cliques of all sizes. 

\textbf{Succinct Clique Trees through Pivoting:} We prove that pivoting can be used
to construct a special data structure called the \emph{Succinct Clique Tree} (SCT).
The SCT stores a unique representation of all cliques, but is much smaller than
the total number of cliques. It can also be built quite efficiently.
Additionally, given the tree, one can easily ``read off" the number of $k$-cliques 
and various local counts in the graph. Remarkably, we can get all counts without storing the entire tree and the storage required at any point is linear in the number of edges. 

%

%
\textbf{Excellent practical performance:} We implement \mainalg{} on a commodity machine.
For global clique counting, \mainalg{} is able to process graphs of up to tens of millions
of edges in \emph{minutes}. Previous results either work only for small values of $k$ (typically up to $10$) or take much longer.
Consider \Fig{timings}, where the time of \mainalg{} is compared
with that of kClist (the state of the art parallel algorithm for clique counting)~\cite{DBS18}. 
In the instances shown kClist did not terminate even after running for 3 days. By
contrast, for the largest {\tt com-orkut} social network with
more than 100M edges, \mainalg{} gets all values of $C_k$ within two hours.
(Typically, in this time, kClist gets $k-$clique counts only up to $k \leq 13$.)

\textbf{Feasible computation of local counts:} \mainalg{} is quite efficient for per-vertex counts, and runs
in at most twice the time for global counts. The times for local clique counting
are given in \Fig{timings}. Even for the extremely challenging problem
of per-edge counts, in most instances \mainalg{} gets these numbers in a few hours.
(For the {\tt com-orkut} social network though, it takes a few days.) 

This allows us to get data shown in \Fig{soc-pokec-occurrences} and \Fig{web-Stanford-occurrences}, 
that plots the frequency distribution
of $k$-cliques. (In other words, for every number $r$, we plot the number of vertices
that participate in $r$ $k$-cliques.) As mentioned earlier, this information is
used for dense subgraph discovery~\cite{SaSePi14,Ts15}. 
To the best of our knowledge, this is the first algorithm that is able to get
such information for real-world graphs. 

\subsection{Related Work} \label{sec:related}

Subgraph counting has an immensely rich history in network science, ranging from applications
across social network analysis, bioinformatics, recommendation systems, graph clustering
(we refer the reader to the tutorial~\cite{SeTi19} and references within). We only
describe work directly relevant to clique counting.

The simplest case of clique counting is \emph{triangle counting}, which has 
received much attention from the data mining and algorithms communities. Recent work
has shown the relevance of counts of large subgraphs (4, 5 vertex patterns)
~\cite{BeHe+11,UganderBK13,SGB16,RKKS17,YiBiKe18}. Local clique counts
have played a significant role in a flurry of work on faster and better algorithms
for dense subgraph discovery and community detection~\cite{SaSePi14,Ts15,BeGlLe16,TPM17}.
The latter results define the ``motif conductance", where cuts are measured by the
number of subgraphs (not just edges) cut. This has been related to higher order
clustering coefficients~\cite{YiBiKe18,YiBiKe19}. These quantities
are computed using local clique counts, underscoring the importance of these numbers.

The problem of counting cliques (and variants such as counting maximal cliques) has received much attention 
both from the applied and theoretical computer science communities~\cite{ChNi85,AlYuZw94,CHKX04,V09}.
Classic techniques like color-coding~\cite{BetzlerBFKN11,ZhWaBu+12} and
path sampling~\cite{SePiKo14,JhSePi15,WaZh+18} have been employed for counting cliques up to size $5$.

For larger cliques, Finocchi-Finocchi-Fusco gave a MapReduce algorithm that uses
orientation and sampling techniques~\cite{FFF15}. Jain and Seshadhri
use methods from extremal combinatorics to give a fast sampling algorithm~\cite{JS17},
that is arguably the fastest approximate clique counter to date. In a remarkable
result, Danisch-Balalau-Sozio gave a parallel implementation (kClist) of a classic algorithm
of Chiba-Nishizeki, which is able to enumerate upto trillions of cliques~\cite{DBS18}.
For exact counting, we consider kClist as the state of the art.
Despite the collection of clever techniques, none of these methods really scale beyond
counting (say) 10-cliques for large graphs.

\textbf{Why local counting is hard:} Note that either parallelism or sampling is used to tame the combinatorial explosion.
Even though (at least for small $k$), one can enumerate all cliques in parallel, local
counting requires updating a potentially global data structure, the list
of all $c_k(v)$ or $c_k(e)$ values. To get the benefits of parallelism, one would 
either have to duplicate a large data structure or combine results
for various threads to get all local counts. While this may be feasible, it
adds an extra memory overhead. 

Sampling methods typically require some overhead for convergence. For local counts,
there are simply too many samples required to get accurate values for (say) all $c_k(v)$ values.
For these reasons, we strongly believe that new ideas were required to get efficient
local counting.

\textbf{Maximal clique enumeration:} Extremely relevant to our approach is a line
of work of maximal clique enumeration. A \emph{maximal clique}
is one that is not contained in a larger clique. Unlike the combinatorial explosion
of $k$-cliques, maximal cliques tend to be much fewer. The first algorithm for this problem
is the classic Bron-Kerbosch backtracking procedure from the 70s~\cite{BK73,A73}.
They also introduced an idea called \emph{pivoting}, that prunes the recursion tree for
efficiency.
Tomita-Tanaka-Takahashi gave the first theoretical analysis of pivoting rules,
and showed asymptotic improvements~\cite{Tomita04}. Eppstein-L\"{o}effler-
Strash
combined these ideas with orientation methods to give a practical and provably fast
algorithm for maximal clique enumeration~\cite{ES11,ELS13}. An important empirical
observation of this line of work is that the underlying recursion tree created with pivoting
is typically small for real-world graphs. This is the starting point 
for our work.

%% file: ideas.tex
\documentclass[../main.tex]{subfiles}

Inspired by the success of maximal clique enumeration through pivoting, 
we design the Succinct Clique Tree (SCT) of a graph for clique counting.

To explain the SCT, it is useful to begin with 
the simple backtracking algorithm
for listing all cliques. For any vertex $v$, let $N(v)$ denote the neighborhood of $v$.
Any clique containing $v$ is formed by adding $v$ to a clique contained in $N(v)$. 
Thus, we can find all cliques by this simple recursive procedure:
for all $v$, recursively enumerate all cliques in $N(v)$. For each such clique,
add $v$ to get a new clique. It is convenient to think of the recursion tree
of this algorithm. Every node of the tree (corresponding to a recursive call)
corresponds to a subset $S \subseteq V$, and the subtree of calls enumerates
all cliques contained in $S$. A call to $S$ makes a recursive call corresponding
to every $s \in S$, which is over the set $N(s) \cap S$ (the neighbors of $v$ in $S$).
We can label every edge of the tree (call them \emph{links} to distinguish from
edges of $G$) with a vertex, whose neighborhood leads to the next recursive call.
It is not hard to see that the link labels, along any path from a root (that
might not end at a leaf), give a clique. Moreover, every clique has such a representation.

Indeed, every permutation of clique forms such a path. A simple and classic
method to eliminate multiple productions of a clique is \emph{acyclic orientations}.
Simply orient the graph as a DAG, and only make recursive calls on out-neighborhoods.
Typically, an orientation is chosen by degeneracy/core decomposition or degree orderings,
so that out-neighborhood sizes are minimized. This is a central technique in all
recent applied algorithms on clique counting~\cite{FFF15,JS17,DBS18}. Yet it
is not feasible to construct the recursion tree to completion, and it is typically truncated
at some depth ($\leq 10$) for large graphs.

Is it possible to somehow ``compress" the tree, and get a unique (easily accessible)
representation of all cliques?


{\bf The power of pivoting:} We discover a suprising answer, 
in pivoting. This was discovered by Bron-Kerbosch in the context 
of \emph{maximal} cliques~\cite{BK73}. 
We describe, at an intuitive level, how it can be applied for global and local clique counting.
For the recursive call at $S$, first pick a \emph{pivot} vertex $p \in S$. 
Observe that the cliques in $S$ can be partitioned into three classes as follows.
For clique $C$ contained in $S$: (i) $p \in C$, (ii) $C \subset N(p)$, (iii)
$C$ contains a non-neighbor of $p$. There is 1-1 correspondence between
cliques of type (i) and (ii), so we could hope to only enumerate
type (ii) cliques. 

Thus, from a recursive call for $S$, we make recursive calls to find cliques
in $N(p) \cap S$, and $N(u) \cap S$ for every \emph{non-neighbor} $u$ of $p$ in $S$.
We avoid making recursive calls corresponding to vertices in $N(p)$. This gives
the main savings over the simple backtracking procedure. The natural choice
of $p$ is the highest degree vertex in the graph induced on $S$. The recursion tree obtained is 
essentially the SCT. We stress that this is quite different from the Bron-Kerbosch recursion
tree. The BK algorithm also maintains a set of excluded vertices since it only
cares for maximal cliques. This excluded set is used to prune away branches that cannot
be maximal; moreover, the pivots in BK are potentially chosen from outside $S$ to increase
pruning. The SCT is constructed in this specific manner to ensure unique clique
representations, which the BK tree does not provide.

%

The SCT is significantly smaller than recursion trees that use
degeneracy orientations (which one cannot feasibly construct).
In practice, it can be constructed efficiently for graphs with tens of millions of edges. 
As before the nodes of the SCT are labeled with subsets (corresponding
to the recursive calls), and links are labeled with vertices (corresponding
to the vertex whose neighborhood is being processed). Abusing notation,
in the following discussion, we refer to a path by the set of link labels in the path.

%

How can we count all cliques using the SCT? Every root to leaf path
in the tree corresponds to a clique, but not all cliques
correspond to paths. This is distinct from the standard recursion tree discussed
earlier, where every clique corresponds to a path from the root. Indeed, this is
why the standard recursion trees (even with degeneracy orientations) are large.

We prove the following remarkable ``unique encoding" property. Within any root to leaf
path $T$, there is a subset of links
$P$ corresponding to the pivot calls. Every clique $C$ in the graph
can be \emph{uniquely} expressed as $(T\setminus P) \cup Q$ for some $Q \subseteq P$
(for a specific path $T$). 
The uniqueness is critical for global and local counting,
since we can simply write down formulas to extract all counts. Thus, the SCT
gives a unique encoding for every clique in the graph. 

Intuitively, the source of compression can be seen in two different ways. The simplest
way is to see that pivoting prunes the tree, because recursive calls
are only made for a subset of vertices. But also, not every clique is
represented by (the link labels of) a path from the root. Thus, there are far fewer paths
in the SCT. 
The final algorithm is quite simple and the main work was coming up with
the above insight. Despite this simplicity, it outperforms even parallel methods 
for exact clique counting by orders of magnitude. 
    
Our main theorem follows. Basically, clique counts can be obtained
in time proportional to the size of the SCT. 
All the technical terms will be formally defined in \Sec{prelims}.

\begin{theorem} \label{thm:main} Let $G$ be an input graph
with $n$ vertices, $m$ edges, and degeneracy $\alpha$. Let $\sct(G)$ be the 
Succinct Clique Tree of graph $G$.

The procedure \mainalg$(G)$ correctly outputs
all global and local counts. For global and per-vertex counts, the running
time is $O(\alpha^2 |\sct(G)| + m + n)$. For per-edge counts, the running time
is $O(\alpha^3 |\sct(G)| + m + n)$. The storage cost is $O(m+n)$.
%
%
\end{theorem}

Empirically, we observe that the SCT is quite small. In the worst-case,
$|\sct(G)|  = O(n 3^{\alpha/3})$, which follows
from arguments by Eppstein-L\"{o}effler-Strash~\cite{ELS13} and Tomita-Tanaka-Takahashi~\cite{Tomita04} (an exponential dependence is necessary because of the NP-hardness of maximum clique). We give a detailed description in \Sec{count}

%% file: algo.tex
\section{Preliminaries} \label{sec:prelims}

We start with the mathematical formalism required to describe the main
algorithm and associated proofs. The input is a simple, undirected
graph $G = (V,E)$, where $|V| = n$ and $|E| = m$. It is convenient to assume
that $G$ is connected. We use vertices
to denote the elements of $V$ (the term \emph{nodes} will be used for a different construct).
We use the following notation for neighborhoods.

\begin{asparaitem}
    \item $N(v)$: This is the neighborhood of $v$.
    \item $N(S,v)$: For any subset of vertices $S$, we use $N(S,v)$ to denote $N(v) \cap S$.
Alternately, this is the neighborhood of $v$ in $S$.
\end{asparaitem}
We will use \emph{degeneracy orderings} (or core decompositions) to reduce
the recursion tree. This is a standard technique for clique counting~\cite{ChNi85,FFF15,JS17,DBS18}.
This ordering is obtained by iteratively removing the minimum degree vertex,
and can be computed in linear time~\cite{MB83}. Typically, one uses this
ordering to convert $G$ into a DAG. The largest \emph{out-degree} is the 
graph degeneracy, denoted $\alpha$. We state this fact as a lemma,
which is considered a classic fact in graph theory and network science.

\begin{lemma} \label{lem:degen}~\cite{MB83} Given a graph $G = (V,E)$, there is a 
linear time algorithm that constructs an ayclic orientation of $G$
such that all outdegrees are at most $\alpha$.
\end{lemma}

The most important construct we design is the \emph{Succinct Clique Tree} (SCT) $\bT$.
The SCT stores special node and link attributes that are key to getting global and local
clique counts, for all values of $k$.
The construction and properties of the SCT are given in the next section. Here, 
we list out technical notation associated with the SCT $\bT$.

Formally, $\bT$ is a tree where nodes are labeled with subsets of $V$, with the following
properties.
\begin{asparaitem}
    \item The root is labeled $V$.
    \item Parent labels are strict supersets of child labels.
    \item Leaves are labeled with the empty set $\emptyset$.
\end{asparaitem}

\medskip

An important aspect of $\bT$ are \emph{link labels}. A link label is a
pair with a vertex of $V$ and a ``call type".
The label is of the form $(v,\mathfrak{p})$ or $(v,\mathfrak{h})$,
where $\mathfrak{p}$ is shorthand for ``pivot" and $\mathfrak{h}$ for ``hold".
For a link label $(v,\cdot)$ of the link $(S,S')$ (where $S \supset S'$ is the parent),
$v$ will be an element of $S$.

Consider a root to leaf path $T$ of $\bT$. We have the following associated
set of vertices. It is convenient to think of $T$ as a set of tree links.

\begin{asparaitem}
    \item $H(T)$: This is the set of vertices associated with ``hold" call types,
    among the links of $T$. Formally, $H(T)$ is
    $\{v | \textrm{$(v,\mathfrak{h})$ is label of link in $T$}\}$.
    \item $P(T)$: This is the set of vertices with ``pivot" calls. Formally $P(T)$ is
    $\{v | \textrm{$(v,\mathfrak{p})$ is label of link in $T$}\}$.
\end{asparaitem}

\medskip

We now describe our algorithm. We stress that the presentation here is different
from the implementation. The following presentation is easier for mathematical
formalization and proving correctness. The implementation is a recursive 
version of the same algorithm, which is more space efficient. This
is explained in the proof of \Thm{main}.

\section{Building the SCT} \label{sec:scr}

We give the algorithm to construct the SCT. 
We keep track
of various attributes to appropriately label the edges.
The algorithm will construct the SCT $\bT$ in a breadth-first manner. Every
time a node is processed, the algorithm creates its children and labels all the new
nodes and links created.

\begin{algorithm}
\caption{\scr($G$) \newline
Output: SCT of $G$ 
}
Find degeneracy orientation of $G$, and let $N^+(v)$ denote the outneighborhood of a vertex $v$.\\
Initialize tree $\bT$ with root labeled $V$.\\
For every $v \in V$, create a child of root with node label $N^+(v)$. Set the 
edge label to $(v,\mathfrak{h})$. \\
Insert all these child nodes into a queue $\bQ$. \\
While $\bQ$ is non-empty: \\
\ \ \ \ Dequeue to get node $\gamma$. Let node label be $S$.\\
\ \ \ \ If $S = \emptyset$, continue. \\
\ \ \ \ Find $p \in S$ with largest $N(S,p)$ value. \label{step:pivot} \\
\ \ \ \ Create child node of $\gamma$ with vertex label $N(S,p)$. Add this node to $\bT$ and set the link
label (of the new link) to $(p,\mathfrak{p})$. Also, add this node to $\bQ$. \label{step:pcall}\\
\ \ \ \ Let $S \setminus (p \cup N(p)) = \{v_1, v_2, \ldots, v_\ell\}$ (listed
in arbitrary order). \\
\ \ \ \ For each $i \leq \ell$: create child node of $\gamma$ labeled
$N(S,v_i) \setminus \{v_1, v_2, \ldots, v_{i-1}\}$. Add this node to $\bT$ and set link label to $(v_i, \mathfrak{h})$.
Also add this node to $\bQ$. \label{step:nncall}\\
Return $\bT$. 
\end{algorithm}

As mentioned earlier, the child of the node labeled $S$ has one child corresponding
to the pivot vertex $p$, and children for all non-neighbors of $p$. Importantly,
we label each ``call" with $\mathfrak{p}$ or $\mathfrak{h}$. This is central to getting
unique representations of all the cliques.

Now for our main theorem about SCT.

\begin{theorem} \label{thm:scr} Every clique $C$ (in $G$) can be \emph{uniquely}
represented as $H(T) \cup Q$, where $Q \subseteq P(T)$ and $T$ is a root to leaf path in $\bT$.
(Meaning, for any other root to leaf path $T' \neq T$, $\forall Q \subseteq P(T')$,
$C \neq H(T') \cup Q$.)
\end{theorem}

We emphasize the significance of this theorem. Every root to leaf
path $T$ represents a clique, given by the vertex set $H(T) \cup P(T)$. Every clique $C$ 
is a subset of potentially many such sets; and there is no obvious bound on this number. So one can think of
$C$ ``occurring" multiple times in the tree $\bT$. But \Thm{scr} asserts that 
if we take the labels into account ($H(T)$ vs $P(T)$),
then there is a \emph{unique} representation or ``single occurrence" of $C$.

\begin{proof} (of \Thm{scr}) 
Consider a node $\gamma$ of $\bT$ labeled $S$. We prove, by induction on $|S|$,
that every clique $C \subseteq S$ can be expressed as $H(T) \cup Q$,
where $T$ is a path from $\gamma$ to a leaf, and $Q \subseteq P(T)$. The theorem
follows by setting $\gamma$ to the root.

The base case is vacuously tree, since for empty $S$, all relevant sets are empty.
Now for the induction. We will have three cases. Let $p$ be the pivot chosen
in \Step{pivot}. (If $S$ is the root, then there is no pivot. We will directly go
to Case (iii) below.)

{\em Case (i): $p \in C$.} By construction, there is a link labeled $(p,\mathfrak{p})$
to a child of $\gamma$. Denote the child $\beta$.
The child $\beta$ has label $N(S,p)$. Observe that $C \setminus p$
is a clique in $N(S,p)$ (since by assumption, $C$ is a clique in $S$.)
By induction, there is a unique representation 
$C\setminus p = H(T) \cup Q$, for path $T$ from the child node
to a leaf and $Q \subseteq P(T)$. Moreover there cannot be a representation
of $C$ by a path rooted at $\beta$, since $N(S,p) \not\ni p$.
Consider the path $T'$ that contains
$T$ and starts from $\gamma$. Note that $H(T') = H(T)$ and $P(T') = P(T) \cup p$.
We can express $C = H(T') \cup (Q \cup p)$, noting that $Q \cup p \subseteq P(T')$.
This proves the existence of a representation. Moreover, there is only one representation
using a path through $\beta$. 

We need to argue that no other path can represent $C$. The pivoting is critical for this step.
Consider any path rooted at $\gamma$, but not passing through $\beta$.
It must pass through some other child, with corresponding links labeled $(v_i,\mathfrak{h})$,
where $v_i$ is a \emph{non-neighbor} of $p$. Since $C \ni p$, a non-neighbor $v_i$
cannot be in $C$. Moreover, for any path $\hat{T}$ passing through these other children,
$\hat{T}$ must contain some non-neighbor. Thus, $\hat{T}$ cannot represent $C$.

{\em Case (ii): $C \subseteq N(S,p)$.} The argument is essentially identical to the one above.
Note that $C\setminus p = C$, and by induction $C\setminus p$ has a unique representation
using a path through $\beta$. For uniqueness, observe that $C$ does not contain a non-neighbor of $p$.
The previous argument goes through as is.

{\em Case (iii): $C$ contains a non-neighbor of $p$.} Recall that $S \setminus (N(p) \cup p)$
(the set of non-neighbors in $S$) is denoted $\{v_1, v_2, \ldots, v_\ell\}$. Let $i$
be the smallest index $i$ such that $v_i \in C$. For any $1 \leq j \leq \ell$, let
$N_j := N(S,v_j) \setminus \{v_1, v_2, \ldots, v_{j-1}\}$. Observe that for all $j$,
there is a child labeled $N_j$. Moreover, all the link labels have $\mathfrak{h}$,
so for path $T$ passing through $N_j$, $H(T) \ni v_j$. Thus, if $T$ can represent $C$,
it cannot pass through $N_j$ for $j < i$. Moreover, if $j > i$, then $N_j \not\ni v_i$
and no path passing through this node can represent $C$.

Hence, if there is a path that can represent $C$, it must pass through $N_i$.
Note that $C\setminus v_i$ is a clique contained in $N_i$. By induction,
there is a unique path $T$ rooted at $N_i$ such that $C \setminus v_i = H(T) \cup Q$,
for $Q \subseteq P(T)$. Let $T'$ be the path that extends $T$ to $\gamma$.
Note that $H(T') = H(T) \cup v_i$, so $C = H(T') \cup Q$. The uniqueness of $T$
implies the uniquesness of $T'$.
%
%
%
%
%
%
%
%
\end{proof}

\section{Getting global and local counts} \label{sec:count}


The tree $\bT$ is succinct and yet one can extract fine-grained information from it about all cliques.

\begin{algorithm}
\caption{\mainalg($G$) \newline
Output: Clique counts of $G$
}
Let $\bT = \scr(G)$.\\
Initialize all clique counts to zero.\\
For every root to leaf path $T$ in $\bT$:\\
\ \ \ \ For every $0 \leq i \leq |P(T)|$, increment $C_{|H(T)|+i}$ by ${P(T) \choose i}$.\\
\ \ \ \ For every $v \in H(T)$ and every $0 \leq i \leq |P(T)|$, increment $c_{|H(T)|+i}(v)$
by ${P(T) \choose i}$. \label{step:ht} \\
\ \ \ \ For every $v \in P(T)$ and every $0 \leq i \leq |P(T)|-1$, increment $c_{|H(T)|+i+1}(v)$
by ${{P(T)-1} \choose i}$. \label{step:pt} \\
\ \ \ \ For every edge $e(u,v), u \in H(T), v \in H(T), u \neq v$ and every $0 \leq i \leq |P(T)|$, increment $c_{|H(T)|+i}(e)$
by ${{P(T)} \choose i}$. \label{step:hht} \\
\ \ \ \ For every edge $e(u,v), u \in P(T), v \in H(T)$ and every $0 \leq i \leq |P(T)|-1$, increment $c_{|H(T)|+i+1}(e)$
by ${{P(T)-1} \choose i}$. \label{step:pht} \\
\ \ \ \ For every edge $e(u,v), u \in P(T), v \in P(T), u \neq v$ and every $0 \leq i \leq |P(T)|-2$, increment $c_{|H(T)|+i+2}(e)$
by ${{P(T)-2} \choose i}$. \label{step:ppt} \\
Output the sets of values $\{C_k\}$, $\{c_k(v)\}$ and $\{c_k(e)\}$.
\end{algorithm}


The storage complexity of the algorithm, as given, is potentially $O(\alpha^2 |\sct(G)|)$,
since this is required to store the tree. In the proof of \Thm{main}, we explain how
to reduce the storage.


\begin{proof} (of \Thm{main}) {\bf Correctness:} By \Thm{scr}, a root to leaf path $T$ of $\bT$ represents exactly
$2^{P(T)}$ different cliques, with ${P(T) \choose i}$ of size $|H(T)| + i$. Moreover,
over all $T$, this accounts for all cliques in the graph. This proves the correctness
of global counts.

Pick a vertex $v \in H(T)$. For every subset of $P(T)$, we get a different
clique containing $v$ (that is uniquely represented by \Thm{scr}). This 
proves the correctness of \Step{ht}. For a vertex $v \in P(T)$,
we look at all subsets containing $v$. Equivalently, we get a different
represented clique containing $v$ for every subset of $P(T)\setminus v$. This 
proves the correctness of \Step{pt}.

Pick an edge $e=(u,v), u \in H(T), v \in H(T)$. For every subset of $P(T)$, we get a different
clique containing $e$ (that is uniquely represented by \Thm{scr}). This 
proves the correctness of \Step{hht}. For an edge $e=(u,v), u \in P(T), v \in H(T)$,
we look at all subsets of $P(T)$ containing $u$. Equivalently, we get a different
represented clique containing $e$ for every subset of $P(T)\setminus u$. This 
proves the correctness of \Step{pht}. For an edge $e=(u,v), u \in P(T), v \in P(T)$,
we look at all subsets of $P(T)$ containing both $u$ and $v$. Equivalently, we get a different
represented clique containing $e$ for every subset of $P(T)\setminus v \setminus u$. This 
proves the correctness of \Step{ppt}.

{\bf Running time (in terms of $|\sct(G)|$):} Consider the procedure $\scr(G)$. Note that the size of $\bT$
is at least $n$, so we can replace any running time dependence on $n$ by $|\bT|$.
The degeneracy orientation can be found in $O(m+n)$~\cite{MB83}. For the actual building
of the tree, the main cost is in determining the pivot and constructing the children
of a node. Suppose a non-root node labeled $S$ is processed. The above mentioned steps can be done
by constructing the subgraph induced on $S$. This can be done in $O(|S|^2)$ time. Since
this is not a root node, $|S| \leq \alpha$ (this is the main utility of the degeneracy
ordering). Thus, the running time of $\scr(G) = O(\alpha^2|\bT|) = O(\alpha^2 |\sct(G)|)$.

Now we look at \mainalg. Note that the subsequent counting steps do \emph{not}
need the node labels in $\bT$; for all path $T$, one only needs $P(T)$ and $H(T)$.
The paths can be looped over by a DFS from the root. For each path,
there are precisely $|P(T)|+1$ updates to global clique counts,
and at most $|H(T) \cup P(T)| \times (|P(T)|+1)$ updates to per-vertex clique counts.
The length of $T$ is at most $\alpha$, and thus both these quantities
are $O(\alpha^2)$. Thus, the total running time is $O(\alpha^2 |\sct(G)|)$ for global and per-vertex clique counting.

Similarly, for each path, at most $|H(T) \cup P(T)|^2 \times (|P(T)|+1)$ updates are made to per-edge clique counts. This quantity is $O(\alpha^3)$. Thus, the total running time is $O(\alpha^3 |\sct(G)|)$.



{\bf Running time (in terms of $n$ and $\alpha$):} One crucial difference between the algorithm of Bron-Kerbosch and \scr{} is that in Bron-Kerbosch, the pivot vertex can be chosen not only from $S$ but also from a set of already processed vertices. Hence, the tree obtained in Bron-Kerbosch can potentially be smaller than that of \mainalg{}. Despite this difference, the recurrence and bound on the worst case running time of $\scr{}$ is the same as Bron-Kerbosch. 

\begin{theorem} \label{thm:main-BK} Worst case running time of \scr{} is $O(n3^{\alpha/3})$. 
\end{theorem}
\begin{proof}
Let $T(s)$ be the worst case running time required by \scr{} to process $S$ where $s=|S|$.

Let $R=S \setminus N(p)$. Let $T_r(s)$ be the worst case running time of processing $S$ when $|R|=r$. Note that when $S$ is being processed it creates a total of $r$ child nodes. 

Thus, $T(s)=\max\limits_r\{T_r(s)\}$. 

Note that all steps other than \Step{pcall} and \Step{nncall} take time $O(s^2)$. Say, they take time $p_1s^2$, where $p_1>0$ is a constant.

Thus, we have that:
\begin{align}
    T_r(s) \leq \sum\limits_{v \in R}{T(|N(S,v)|) 
    + p_1s^2}.
\end{align}

Moreover, 
\begin{align}
    |N(S,v)| \leq s-r \leq s-1,\forall v \in R. 
\end{align}

This is because $p$ has the largest neighborhood in $S$ and $p$'s neighborhood is of size atmost $s-r$, and since $|S|\geq 1, s-r \leq s-1$.

Thus, Lemma 2 and Theorem 3 from ~\cite{Tomita04} hold, which implies that $T(s)=O(3^{s/3})$. Since there are $n$ vertices and their outdegree is atmost $\alpha$, the worst case running time of \scr{} (which is also an upper bound for $|sct(G)|$) is $nT(\alpha)=O(n3^{\alpha/3})$ and hence, worst case running times of \mainalg{} for obtaining global, per-vertex and per-edge clique counts are $O(n\alpha 3^{\alpha/3})$, $O(n\alpha^2 3^{\alpha/3})$ and $O(n\alpha^3 3^{\alpha/3})$, respectively.



\end{proof}



{\bf Storage cost:} Currently, \mainalg{} is represented through
two parts: the construction of $\sct(G)$ and then processing it to get
clique counts. Conceptually, this is cleaner to think about and it makes the proof
transparent. On the other hand, it requires storing $\sct(G)$, which is potentially
larger than the input graph. A more space efficient implementation is obtained
by combining these steps. 

We do not give full pseudocode, since it is somewhat
of a distraction. (The details can be found in the code.) Essentially,
instead of constructing $\sct(G)$ completely in breadth-first manner,
we construct it depth-first through recursion. This will loop
over all the paths of $\bT$, but only store a single path at any stage.
The updates to the clique counts are done as soon as any root to leaf
path is constructed. The total storage of a path is the storage
for all the labels on a path. As mentioned earlier in the
proof of \Thm{main}, all non-root
nodes are labeled with sets of size at most $\alpha$. The length
of the path is at most $\alpha$, so the total storage is $O(\alpha^2)$.
A classic bound on the degeneracy is $\alpha \leq \sqrt{2m}$ (Lemma 1 of~\cite{ChNi85}),
so the storage, including the input, is $O(m+n)$.

\end{proof}

{\bf Parallel version of \mainalg:} While this is not central to our results,
we can easily implement a parallel version of \mainalg{} for \emph{global} clique counts.
We stress that our aim was not to delve into complicated parallel algorithms,
and merely to see if there was a way to parallelize the counting involving minimal code changes.
The idea is simple, and is an easier variant of the parallelism in kCList~\cite{DBS18}. 
Observe that the children of the root
of $\sct(G)$ correspond to finding cliques in the sets $N^+(v)$, for all $v$.
Clique counting in each of these sets can be treated as an independent problem,
and can be handled by an independent thread/subprocess. Each subprocess
maintains its own array of global clique counts. The final result
aggregates all the clique counts. The change ends up being a few lines of code
to the original implementation.

Note that this becomes tricky for local counts. Each subprocess cannot
afford (storage-wise) to store an entire copy of the local count data structure.
The aggregation step would be more challenging. Nonetheless, it should be feasible
for each subprocess to create local counts for $N^+(v)$, and appropriately
aggregate all counts. We leave this for future work.

{\bf Counting $k$-cliques for a specific $k$:} \mainalg{} can be modified to obtain clique counts upto a certain user specified $k$ (instead of counting for all $k$). Whenever the number of links marked $\mathfrak{h}$ becomes greater than $k$ in any branch of the computation, we simply truncate the branch (as further calls in the branch will only yield cliques of larger sizes).

%% file: results.tex
{\bf Preliminaries:} All code for \mainalg{} is available here: \href{https://bitbucket.org/sjain12/pivoter/}{https://bitbucket.org/sjain12/pivoter/}. We implemented our algorithms in C
and ran our experiments on a commodity machine equipped
with a 1.4GHz AMD Opteron(TM) processor 6272 with 8 cores and
2048KB L2 cache (per core), 6144KB L3 cache, and 128GB
memory.
We performed our experiments on a collection of social networks, web networks, and
infrastructure networks 
from SNAP [49]. 
The graphs are simple and undirected (for graphs that are directed, we ignore the direction).
A number of these graphs have more than 10 million edges, and the largest has more than
100 million edges.  Basic properties of these graphs are presented in \Tab{main}. 

The data sets are split into two parts, in \Tab{main}. The upper part are
instances feasibly solved with past work (notably kClist40~\cite{DBS18}), while
the lower part has instances that cannot be solved with previous algorithm (even
after days). We give more details in \Sec{time}.

{\bf Competing algorithms:} We compare with (what we consider) are the state
of the art clique counting algorithms: Tur\'{a}n-Shadow (TS)~\cite{JS17} and kClist40~\cite{DBS18}.

\textbf{kClist40:} This algorithm by Danisch-Balalau-Sozio~\cite{DBS18} uses 
degeneracy orientations and parallelization to enumerate all cliques. 
The kClist40 algorithm, to the best of our knowledge, is the only existing algorithm
that can feasibly compute all global counts for some graphs. Hence, our main focus
is runtime comparisons with kClist40.

We note that the implementation of kClist40 visits every clique, but only
updates the (appropriate) $C_k$. While it could technically compute local counts,
that would require more expensive data structure updates. Furthermore, there would
be overhead in combining the counts for independent threads, and it is not
immediately obvious how to distribute the underlying data structure storing
local counts. As a result, we are
unaware of any algorithm that computes local counts (at the scale of dataset in \Tab{main}).

We perform a simple optimization of kClist40, to make counting faster.
Currently, when kClist40 encounters a clique, it enumerates every smaller clique contained inside it. 
For the purpose of counting though, one can trivially count all subcliques
of a clique using formulas. We perform this optimization (to have a fair comparison with kClist40),
and note significant improvements in running time.

In all our runs, for consistency, we run kClist with 40 threads. 
Note that we compare the \emph{sequential} \mainalg{} with the \emph{parallel} kClist40.

\textbf{TS:} This is an approximate clique counting algorithm for $k$ upto $10$ ~\cite{JS17}. 
It mines dense subgraphs (shadows) and samples cliques within the dense subgraphs
to give an estimate. For fast randomized estimates, it is arguably the fastest algorithm.
It runs significantly faster than a sequential implementation of kClist, but is typically comparable
with a parallel implementation of kClist.
It requires the entire shadow to be available for sampling which can require considerable space. 

\subsection{Running time and comparison with other algorithms} \label{sec:time}

\textbf{Running time for global counting:} We show the running time results in \Tab{main}.
For most of the graphs, \mainalg{} was able to count all $k$-cliques in seconds or minutes.
For the largest {\tt com-orkut} graph, \mainalg{} ran in 1.5 hours. 
This is a huge improvement on the state of the art. For the ``infeasible"
instances in \Tab{main}, we do not get results even in two days using previous algorithms.
(This is consistent with results in Table 2 of~\cite{DBS18}, where some of the graphs are also
listed as ``very large graphs" for which clique counting is hard.)

A notable hard instance is {\tt com-lj} where \mainalg{} is unable to get all
clique counts in a day. Again, previous work also notes this challenge,
and only gives counts of $7$-cliques. We can get some partial results for {\tt com-lj},
as explained later.

%

\begin{table*}[t]
\centering
\begin{adjustbox}{max width=\textwidth}
\begin{tabular}{|c|c|c|c|c|c|c|c|c|}
\hline
\textbf{Graph}         & \textbf{Vertices}     & \textbf{Edges}      & \textbf{Degen}  & \textbf{Max clique}
& \multicolumn{1}{|p{1.2cm}|}{\centering \mainalg{} \\ {\bf ($C_k$)}}
& \multicolumn{1}{|p{1.4cm}|}{\centering \mainalg{} \\ {\bf ($c_k(v)$)}}
& \multicolumn{1}{|p{1.4cm}|}{\centering \mainalg{} \\ {\bf ($c_k(e)$)}}
& \multicolumn{1}{|p{1.2cm}|}{\centering \mainalg{} \\ {\bf ($C_k$ parallel)}} \\
\hline
\multicolumn{9}{|c|}{Feasible by previous algorithms} \\
\hline
dblp-v5 & 1.56E+06 & 2.08E+06 & 15 & 10 & 7 & 7 & 8 & 19 \\
dblp-v7 & 3.67E+06 & 4.18E+06 & 19 & 12 & 15 & 16 & 19 & 34 \\
amazon0601 & 4.03E+05  &  2.44E+06   & 10  &11&  4  & 5  & 6 & 4 \\
web-Google & 8.76E+05  &  4.32E+06   & 44 & 44 & 8 & 9 &15 & 9\\
youtube &1.13E+06   & 2.99E+06  &  51  &17  &7 & 8  &11 & 9 \\
cit-Patents &3.77E+06    &1.65E+07   & 64 & 11 & 40  &41 & 53 & 46 \\
soc-pokec  & 1.63E+06  &  2.23E+07  &  47 & 29 & 68& 75 &93 & 44 \\
\hline
\multicolumn{9}{|c|}{Not feasible for previous algorithms} \\
\hline
Stanford  &  2.82E+05   & 1.99E+06  &  71 &  61 & 5 &  5 & 38 & 3 \\
BerkStan  &  6.85E+05  &  6.65E+06   & 201 &201 &25 & 26  &237  & 9\\
as-skitter  & 1.70E+06   & 1.11E+07 &   111 & 67  & 120 & 200 &9245  & 75\\
com-orkut  & 3.07E+06  &  1.17E+08  &  253 & 51 & 5174   & 8802  & 99389  & 3441 \\ \hline
com-lj  & 4.00E+06  &  3.47E+07  &  360 & - & -   & -  & -  & 108000* \\ 
\hline
\end{tabular}
\end{adjustbox}
\caption{Table shows the sizes, degeneracy, maximum clique size, and the time taken (in seconds) by \mainalg{} to obtain global $k-$clique counts, per-vertex and per edge $k-$cliques counts for all k. *For the com-lj graph, we were not able to get all $k-$clique counts in 1 day so we tested for the maximum $k$ we could count in about a day. \mainalg{} was able to count the number of 9-cliques in 30 hours whereas kClist40 had not terminated even after 6 days.}
\label{tab:main}
\end{table*}

\begin{table}[]
\centering
\begin{adjustbox}{max width=\textwidth}
\begin{tabular}{|l|r|r|r|}
\hline
\textbf{Graph}         & \textbf{k=13,TS}   & \textbf{k=13, kClist40}  & \textbf{all $k$, \mainalg{}} \\ 
\hline
Stanford  & 230 & 12600 &5 \\ 
BerkStan  & 1198 &   > 172800 &25 \\ 
as-skitter & 798 & 12480 &120 \\ 
com-orkut  & > 28800 &  > 172800  &5174\\ 
\hline
\end{tabular}
\end{adjustbox}
\caption{Time taken in seconds by the state-of-the-art randomized (TS, short for Tur\'anShadow) and parallel (kClist40) algorithms. Note that \mainalg{} obtains \textbf{all} $k-$clique counts for these graphs in a fraction of the time taken by other methods to count just 13-cliques.} 
\label{tab:comparison}
\end{table}

\begin{table}[]
\centering
\begin{adjustbox}{max width=\textwidth}
\begin{tabular}{|l|r|r|r|}
\hline
\textbf{k}          & \textbf{$k$-cliques} &  \textbf{kClist40}  & \textbf{\mainalg{}} \\ 
\hline
7 & 4.49E+15 & 2.2 hours & 1.2 hours \\
8  & 1.69E+16 & 42.5 hours & 6.4 hours \\ 
9 &  5.87E+17 & > 6 days & 30 hours \\ 
10  & 1.89E+19 & > 6 days & 5.9 days \\ 
\hline
\end{tabular}
\end{adjustbox}
\caption{Table shows the time taken to count $k$-cliques for com-lj graph. For $k$=9, \mainalg{} terminated in about 30 hours where kClist40 had not terminated in 6 days.} 
\label{tab:com-lj}
\end{table}

\begin{figure*}[t!]
    \begin{subfigure}[b]{0.33\textwidth}
    \centering
    \includegraphics[width=\textwidth]{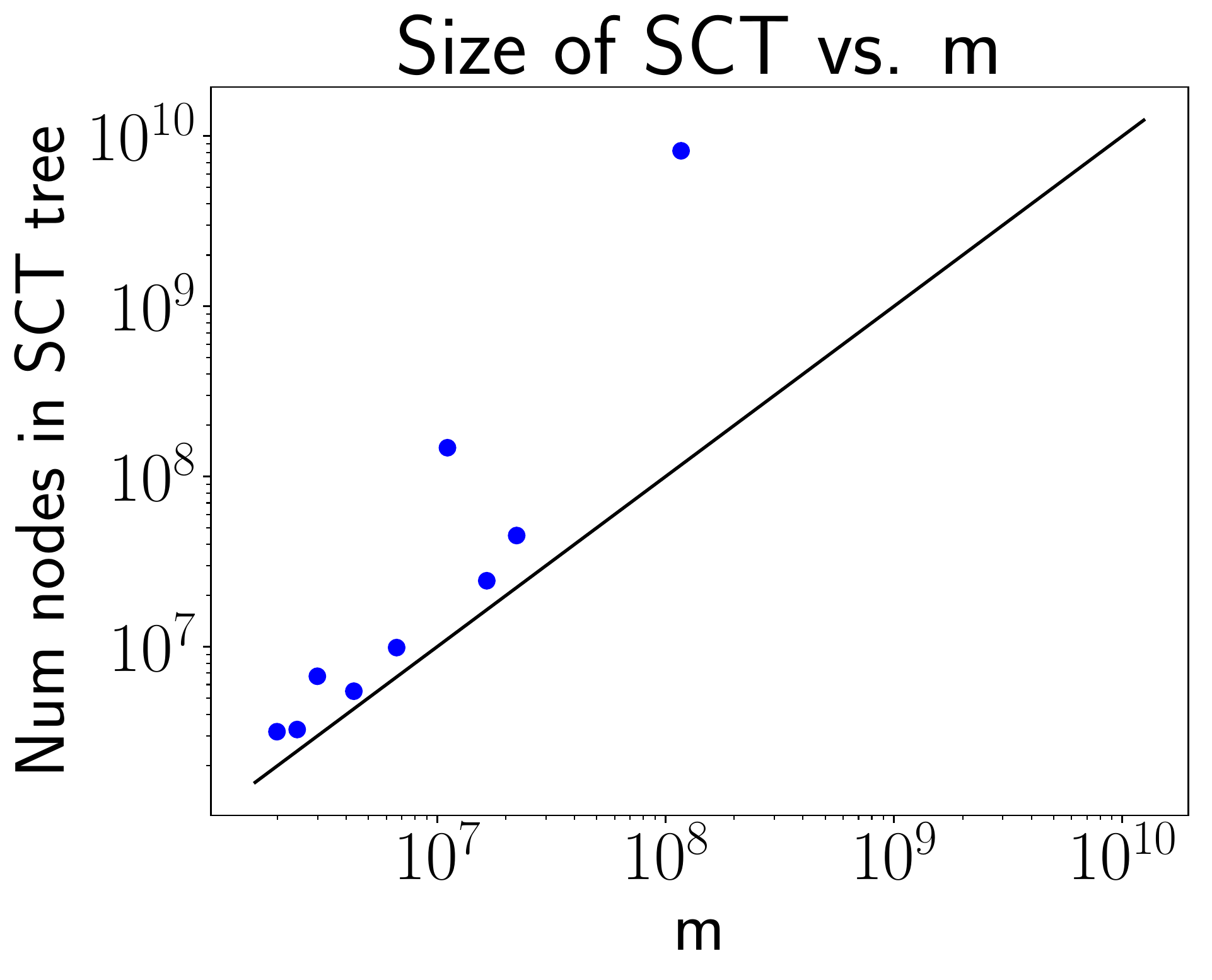}
        \caption{Number of nodes in SCT vs m}
        \label{fig:treesize}
    \end{subfigure}
~
\begin{subfigure}[b]{0.33\textwidth}
    \centering
    \includegraphics[width=\textwidth]{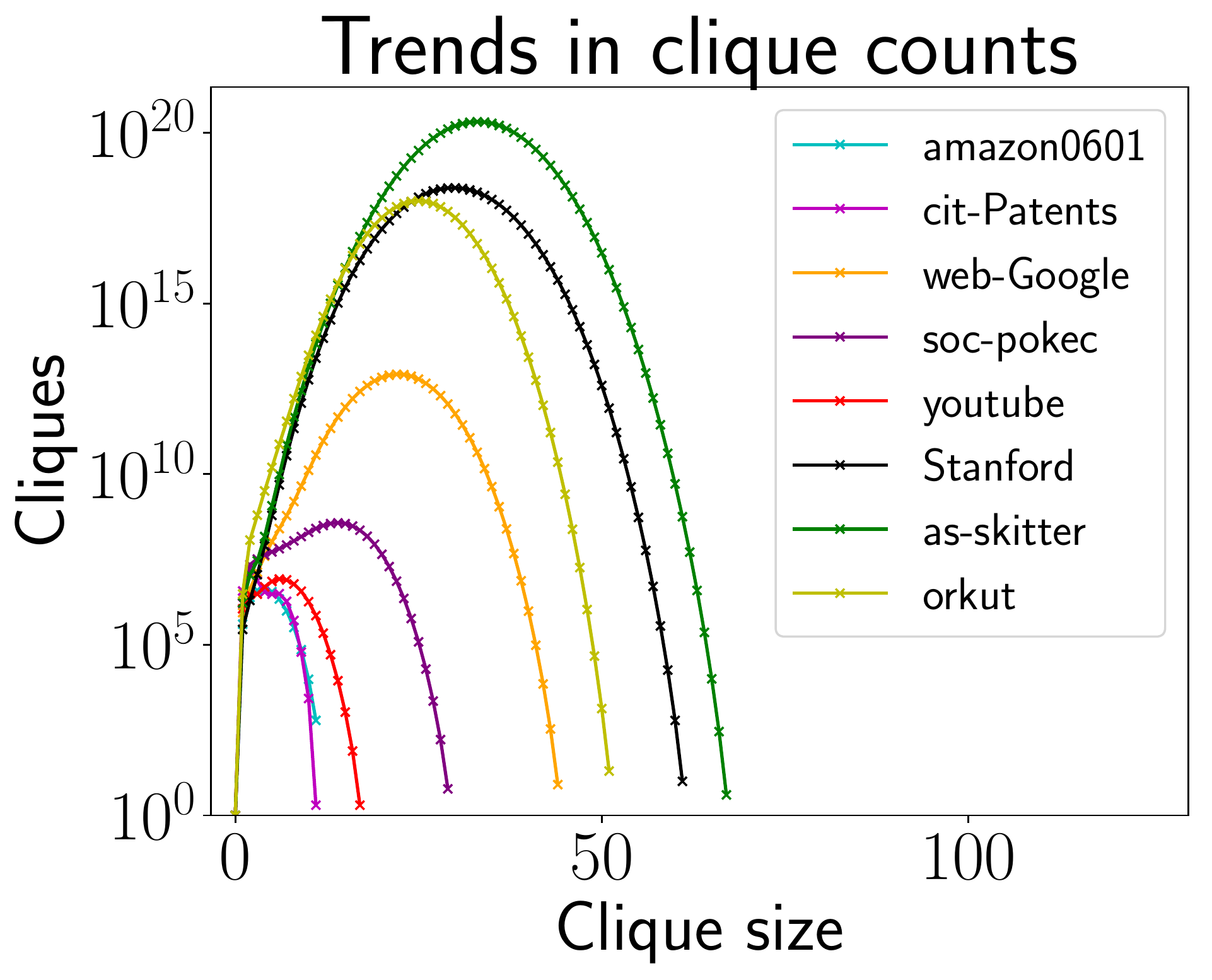}
    \caption{Trends in different graphs}
    \label{fig:trends}
    \end{subfigure}
~
    \begin{subfigure}[b]{0.33\textwidth}
    \centering
    \includegraphics[width=\textwidth]{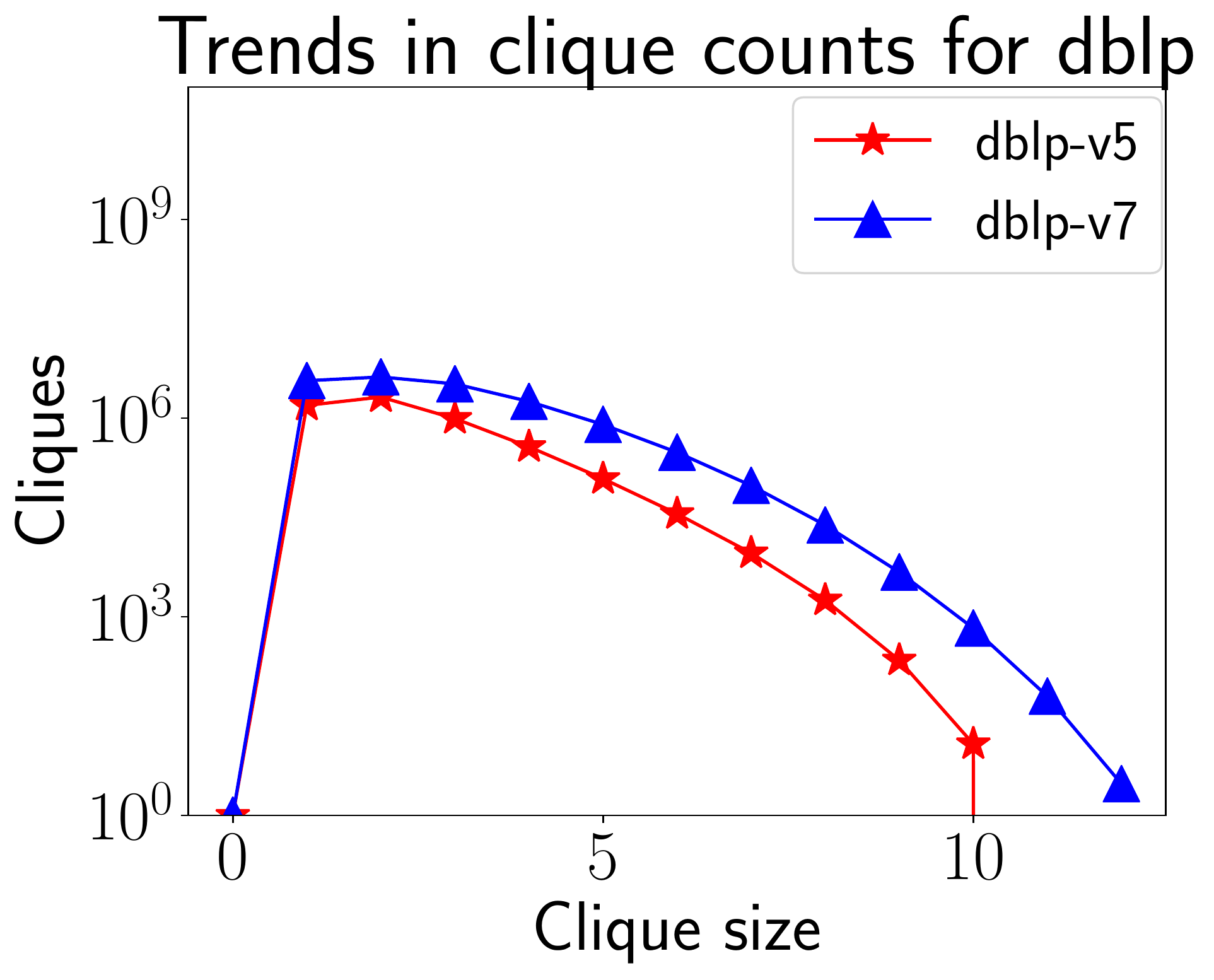}{}
    \caption{Trends in dblp over time.}
    \label{fig:trends-dblp}
    \end{subfigure}
     
\caption{\Fig{treesize} shows the number of nodes in the SCT vs the number of edges (m) for different graphs. The running time of \mainalg{} is directly proportional to the SCT size which seems to be roughly linear in the number of edges. \Fig{trends} shows the trends in clique counts for a number of graphs. For some of the graphs, the complete distribution of their clique counts has been obtained for the first time. \Fig{trends-dblp} shows the trends in the clique counts of 2 different versions over time of the dblp graph. }

\end{figure*}


\textbf{Feasible local counting:} Notably, \mainalg{} can get per-vertex
counts in less than twice the time of global clique counting. Thus, we get 
results for more graphs in a few minutes, and can process the {\tt com-orkut}
graph within 3 hours. We consider this a significant achievement, given
the combinatorial explosion of clique counting.

\mainalg{} is also able to get per-edge clique counts, though
it can take an order of magnitude more time than global clique counting. Note that for obtaining the per-vertex and per-edge $k-$clique counts, the result data structure can become extremely large. Indeed, most of the time is spent in updating the data structure, rather than in constructing the SCT.
Nonetheless, for all but the {\tt as-skitter} and {\tt com-orkut} graph,
it runs in minutes. 

\textbf{Comparison with state of the art:} We only focus on the ``infeasible"
instances of \Tab{main}. For all the other instances, both \mainalg{} and kClist40
get results within two minutes. For space considerations, we do not report all
the running times for such instances. It is worth noting that the sequential
\mainalg{} is comparable to the parallel kClist40 (when they both terminate).

In \Tab{comparison}, we report times on TS and kClist40 on the hard datasets.
We are unable to get all values of $C_k$ using either of these two method. We run
these algorithms for up to 100 times the running time of \mainalg{} or two days, whichever
is shorter. We try to count the largest feasible clique count.

Let us focus on kClist40, where we cannot go beyond counting 13-cliques
(we note that this is consistent with results reported in~\cite{DBS18}). 
Notably, in the {\tt BerkStan} graph, kClist40 needs more than 2 days to count
13-cliques, while \mainalg{} gets all clique counts in a minute. As mentioned
earlier, clique counting on the large {\tt com-orkut} graph is done in a few
hours by \mainalg, while even counting 13-cliques takes kClist40 more than two days.

TS also does not scale well for larger cliques and \mainalg{} is faster than TS. For example, for the {\tt Stanford} graph, TS required 230 seconds to estimate the number of 13-cliques whereas \mainalg{} obtained all $k-$clique counts in 5 seconds. Similar trends are observed with other graphs.

\textbf{Parallel global clique counting:} As mentioned in \Sec{count}, we do a simple
parallelization of the global clique counting of \mainalg{} using 30 threads. It gives moderate benefits
for most instances, and about a factor two speedup for large instances.
For the challenging {\tt com-lj} instances, the effect is much more dramatic. 
We are able to count $7$-cliques in an hour using the parallel \mainalg, while
the sequential version takes more than a day.

\textbf{Performance on com-lj.} This is a particularly challenging graph. 
The sequential version of \mainalg{} for counting all $k$-cliques did not terminate within a day, 
so we used the parallel version of our algorithm to show a comparison for global
counts upto $k=10$. We can truncate the SCT to get cliques of some fixed size.
\Tab{com-lj} shows the results. Even for this graph, the parallel version of \mainalg{} is faster than kClist40 for $k=7$ and beyond. kClist40 did not terminate after six days,
for $k = 9$ and beyond. We note the astronomical number
of 10-cliques ($> 10^{19}$), which makes enumeration infeasible, but \mainalg{}
was able to get the exact count.


\textbf{Size of $\sct(G)$:} In \Fig{treesize}, we plot
the number of nodes of $\sct(G)$ as a function of the number
of edges in $G$. We observe that for most graphs, the size
is quite close to $m$, explaining why \mainalg{} is efficient.

\subsection{Demonstrations of \mainalg} \label{sec:appl}

Global and local cliques have numerous applications. It is 
outside the scope of this work for detailed demonstrations, but
we show a few examples in this section.

As mentioned earlier, local clique counts are an important aspect
of graph processing. In \Fig{soc-pokec-occurrences} and \Fig{web-Stanford-occurrences}, we plot the per-vertex
clique distributions, also called the \emph{graphlet} degree
distribution in bioinformatics~\cite{Pr07} for the as-skitter and web-Stanford graphs. We choose values
of $k = 5, 10, 15, 20, 25$. Then, we plot the function $f_k(b)$
that is the number of vertices that participate in $b$ $k$-cliques.
We notice interesting trends. While the as-skitter graph
has a nicely decaying $f_k$ function, there is much more noise
in web-Stanford. It would be interesting to design models
that can capture such behavior in the local clique counts.


In \Fig{trends}, we plot the $C_k$ values for a number of graphs.
We notice, for example, that the {\tt soc-pokec} network has a ``flatter" distribution of $C_k$ for some of the initial values,
while the {\tt com-orkut} graph looks much closer to a binomial distribution. The latter suggests that the bulk of cliques are coming
from the maximum clique in the {\tt com-orkut} graph, but not
so in the {\tt soc-pokec} graph.

In \Fig{trends-dblp}, we plot the $k$-clique counts (vs $k$)
for two different versions across time for the DBLP citation network~\cite{Aminer}. 
Interestingly, despite the later version only having less than twice as many edges, 
the clique distribution (plotted in semilog) has a much bigger difference. It appears
that the graph is becoming significantly dense in certain part.
This sort of analysis may help in understanding dynamic graphs.

\section{Future work} \label{sec:future}

We provide an exact clique counting algorithm that counts all $k$-cliques in a fraction of the time of other state-of-the-art parallel algorithms. One of the key ideas is the use of pivoting to create
the SCT, and succinct representation of all the cliques of the
graph. The success of~\cite{DBS18} in using parallelization
for clique counting suggests combining their ideas with our pivoting
techniques. We may be able to come up with an efficient parallel
building of the SCT that is much faster than our current implementation. Indeed, the results on the {\tt com-lj} graph suggest
that even \mainalg{} has its limits for real data.

An orthogonal approach would be to exploit the sampling techniques
in the Tur\'{a}n-Shadow algorithm~\cite{JS17}. For many subgraph
counting problems, randomization has been the key to truly practical
algorithms. We believe that \mainalg{} could be made faster with 
these ideas.

Moreover, it also gives per-edge and per-vertex $k-$clique counts. This is the first time that $k-$clique counts are known for many of the graphs we experimented with and this will open doors for further use of cliques in generation and analysis of graphs.